\documentclass[twocolumn,showpacs,preprintnumbers,amsmath,amssymb]{revtex4}

\usepackage{graphicx}
\usepackage{dcolumn}
\usepackage{bm}
\usepackage{color}

\begin{document}

\preprint{APS/*****}

\title{Hall effect of spin-chirality origin in a triangular-lattice helimagnet Fe$_{1.3}$Sb}

\author{Y. Shiomi$^{\, 1}$}
\author{M. Mochizuki$^{\, 1}$}
\author{Y. Kaneko$^{\, 2}$}
\author{Y. Tokura$^{\, 1,2,3}$}
\affiliation{$^{1}$
Department of Applied Physics, University of Tokyo, Tokyo 113-8656, Japan }
\affiliation{$^{2}$
Multiferroics Project, ERATO, Japan Science and Technology Agency (JST), Tokyo 113-8656, Japan
}
\affiliation{$^{3}$
Cross-Correlated materials Research Group (CMRG)/ Correlated Electron Research Group (CERG), ASI, RIKEN, Wako 351-0198, Japan
}

\date{\today}

\begin{abstract}
 We report on a topological Hall effect possibly induced by scalar spin chirality in a quasi-two-dimensional helimagnet Fe$_{1+ \delta}$Sb. In the low-temperature region where the spins on interstitial-Fe (concentration $\delta \sim 0.3$) intervening the $120^\circ$ spin-ordered triangular planes tend to freeze, a non-trivial component of Hall resistivity with opposite sign of the conventional anomalous Hall term is observed under magnetic field applied perpendicular to the triangular-lattice plane. The observed unconventional Hall effect is ascribed to the scalar spin chirality arising from the heptamer spin-clusters around the interstitial-Fe sites, which can be induced by the spin modulation by the Dzyaloshinsky-Moriya interaction.       
\end{abstract}

\maketitle


The interplay between electron transport and magnetism has been a central subject of condensed-matter physics. The Hall effect (HE) in ferromagnets as one such example is known to have two contributions; one is the ordinary Hall effect induced by Lorentz force, which is proportional to magnetic field ($H$), and the other is the anomalous Hall effect (AHE), which arises usually from the spin-orbit interaction and is driven by the magnetization ($M$). While AHE was discovered more than a century ago, the theoretical elucidation has been a controversial issue \cite{review}. Recently, a connection to Berry phase of Bloch electrons in $\bm{k}$-space was proposed as the origin of intrinsic AHE \cite{jungwirth, Monoda}, which has turned out to be a non-perturbative extension of Kaplus-Luttinger theory \cite{KL}, and to successfully explain the {\it intrinsic} AHE observed in some of ferromagnets, such as Fe \cite{Yao} and SrRuO$_{3}$ \cite{Fang}. In this theory, the energy gap induced by the spin-orbit interaction at the band-crossing point acts as a fictitious magnetic field (Berry-phase curvature) in ${\bm k}$-space and hence ensures the non-dissipative nature of anomalous Hall current. \par

While the anomalous Hall resistivity is usually proportional to $M$, a counter example has recently been found in the case of a ferromagnetic pyrochlore Nd$_{2}$Mo$_{2}$O$_{7}$ \cite{Taguchi}; the magnitude of the Hall resistivity increases with decreasing temperature ($T$) and the $H$-dependence seems not to be related straightforwardly to that of $M$. This HE has been interpreted as originating from the noncoplanar configuration of Mo spins. The scalar spin chirality [$\chi_{ijk} = \bm{S}_{i} \cdot (\bm{S}_{j} \times \bm{S}_{k})$] associated with the noncoplanar spin-configuration endows the conduction electron with the Berry phase in ${\bm k}$-space, just as the relativistic spin-orbit interaction does. A similar unconventional HE was also reported in another pyrochlore Pr$_{2}$Ir$_{2}$O$_{7}$ \cite{nakatsuji}. \par

Although the noncoplanar spin configuration is realized in helical magnets under applied $H$, the HE of the scalar-spin-chirality origin is usually not observed because the total scalar spin chirality summed over the whole lattice sites often becomes zero 
\cite{chirality-note}. In the case of the triangular lattice with three-sublattice spin order ({\it e.g.} $120^\circ$-spin structure), the whole scalar spin chirality cancels out and the contribution to HE is not expected \cite{Martin, Akagi}. Recently, however, in a triangular-lattice antiferromagnet PdCrO$_{2}$ where Cr$^{3+}$ spins order with a $120 ^\circ$-structure below $T_{N} =37$ K, an unconventional HE was observed under $H$ applied parallel to the $c$ axis below $T^{*} \sim 20$ K that is noticeably lower than $T_{N}$ \cite{Maeno}. It was speculated in [\onlinecite{Maeno}] that the spin structure would change so as to have a finite scalar spin chirality below $T^{*}$ under $H$ applied parallel to the spin-spiral plane, although such a subtle spin-structural modification could not be proved. In this Letter, we report on a more explicit case of the spin-chirality induced HE on the triangular-lattice magnet Fe$_{1+ \delta}$Sb, in which we could identify that Dzyaloshinsky-Moriya (DM) interaction modifies the spin structure in the spin clusters associated with the interstitial-Fe spins to generate the net scalar spin chirality. The DM-interaction mediated topological Hall effect (THE) as proposed here may be found in many of triangular-lattice magnets with modified $120^\circ$-spin structures.  \par 



\par

The crystal structure of Fe$_{1+ \delta}$Sb (NiAs-type) is depicted in Fig. 1(c). Fe and Sb triangular nets are alternately stacked along the $c$ axis, while hosting the interstitial-Fe (Fe(i)) atoms (content $\delta$). The dotted circles in Fig. 1(c) indicate the possible positions which can be randomly occupied by the Fe(i) atoms; these positions correspond to the apex connecting the upper and lower adjacent-lattice Fe-triads. Fe$_{1+\delta}$Sb is always off-stoichiometric ($\delta \neq 0$) and is known to be stabilized for $0.08 < \delta <0.35$ \cite{Kumar}. Spins of lattice-Fe (Fe(l)) order in a $120 ^\circ$-spin structure within the $c$ plane and ferromagnetically along the $c$ axis below $T_{N}$ (Fig. 1(d)), as determined by the neutron diffraction measurement for Fe$_{1.14}$Sb \cite{yashiro}. The $T_{N}$ value obtained from the M\"ossbauer spectroscopy and susceptibility data decreases from $200$ K to $50$ K with increasing Fe(i) content $\delta$ \cite{Kumar}. Spins of Fe(i) do not order at $T_{N}$ but freeze in the much lower-$T$ region than $T_{N}$ with the thermal hysteresis in $M$-$T$ curve (see Fig. 1(b)) \cite{Amor}. According to the powder neutron diffraction measurement \cite{yashiro}, the $120^\circ$-spin structure of the Fe(l) spins is still retained in such a low-$T$ region independently of the spin glass transition of the Fe(i) spins. It is anticipated that the long range $120^\circ$-spin order and the spin-glass (micto-glass) state around Fe(i) coexist at the lowest $T$ \cite{Amor}. Although the magnetic properties have been intensively investigated for Fe$_{1+\delta}$Sb as described above, little attention has been paid to the transport properties. The development of the local magnetic order on Fe(i) sites is expected to affect the net scalar spin chirality of the system which would be totally canceled out in the $120^\circ$-spin structure without Fe(i). We show here that the topological Hall resistivity as induced by the scalar spin chirality is observed in external $H$ applied parallel to the $c$ axis in the low-$T$ region where the chiral magnetic clusters are built up around the Fe(i) sites.  
  \par      


Single crystals of Fe$_{1.3}$Sb ($\delta =0.3$) were grown by a Bridgman method,
 following the procedure described in literature \cite{yashiro}. The $T$- and $H$-dependences of $M$ were measured by SQUID and extraction-type magnetometers. The Hall resistivity and magnetoresistance were measured within the $c$ plane under $H$ applied parallel to the $c$ axis. \par

\begin{figure}[hbtp]
\begin{center}
\includegraphics[width=8cm]{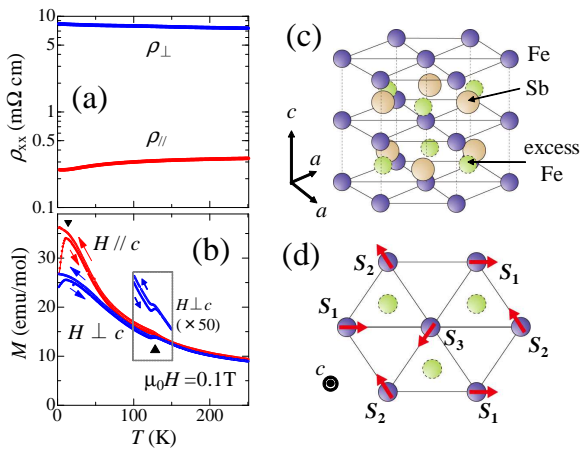}
\caption{(Color online.) (a) Temperature dependence of the in-plane and out-of-plane resistivities, $\rho_{||}$ and $\rho_{\perp}$, for Fe$_{1.3}$Sb. (b) $T$ dependence of the magnetization ($M$) in external magnetic field ($H$) parallel or perpendicular to the $c$ axis. Two triangles indicate the ordering temperatures of the in-plane Fe(l) spins ($T_{N}$) and the interstitial-Fe(i) spins, respectively. The inset shows the magnified $M$-$T$-curve around $T_{N}$ in $H \perp c$. (c) Schematic view of the crystal structure of Fe$_{1+ \delta}$Sb. Fe(i) atoms, whose concentration is $\delta$, randomly occupy the sites shown by dotted circles. (d) Top view of the Fe(l) and Fe(i) sites where the (red) arrows indicate the directions of the Fe(l) spins in the $120^\circ$-spin structure below $T_{N}$.} 
\label{fig1}
\end{center}
\end{figure}

 Figure 1(a) shows the $T$ dependence of resistivity measured within and out of the $c$ plane ($\rho_{||}$ and $\rho_{\perp}$ respectively). $\rho_{||}$ slightly decreases with decreasing $T$, but the residual resistance ratio to the room-temperature value is less than $2$, suggesting the appreciable effect of disorder by randomly occupying Fe(i). On the other hand, $\rho_{\perp}$ is one order of magnitude larger than $\rho_{||}$ and slightly increases with decreasing $T$, indicating the quasi-two-dimensional transport characteristic of the layered lattice structure shown in Fig. 1(c). We show in Fig. 1(b) the $T$ dependence of $M$ measured in $H$ parallel or perpendicular to the $c$ axis. A transition is observed around $130$ K, below which Fe(l) spins order in the in-plane $120^\circ$ spin structure (Fig. 1(d)) \cite{picone}. According to the previous study \cite{Kumar}, the $T_{N}$ value corresponds to that of $\delta \sim 0.25$, which is slightly smaller than the nominal value ($\delta \sim 0.3$) in the present crystal. The decrease of $M$ upon $T_{N}$ in $H \perp c$ is larger than that in $H || c$. The observed anisotropy is consistent with the magnetic transition to the in-plane $120^\circ$-spin order. Since the Fe(i) spins do not order magnetically at $T_{N}$, the Curie-like increase of $M$ persists below $T_{N}$ as observed. Around $10$ K, $M$ shows the maximum accompanying the large thermal hysteresis both for $H||c$ and $H\perp c$. In the present sample, the spin glass transition relevant to the Fe(i) spins seems to occur around $10$ K, which is lower than the value reported in literature ($\sim 30$ K) \cite{yashiro, Amor}. 
\par

\begin{figure}[hbtp]
\begin{center}
\includegraphics[width=8cm]{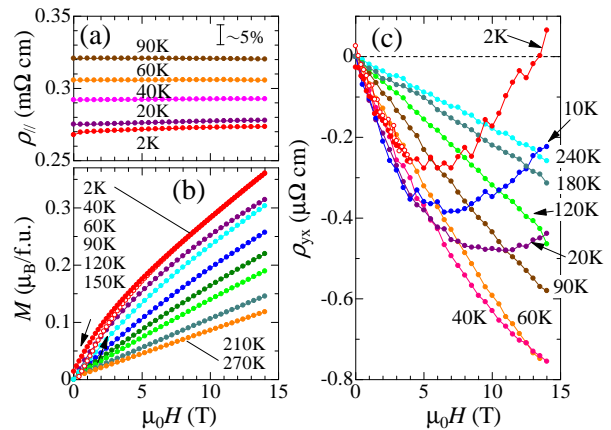}
\caption{(Color online.) Magnetic-field ($H$) dependence of (a) the in-plane resistivity ($\rho_{||}$), (b) the magnetization ($M$), and (c) the Hall resistivity $\rho_{yx}$ under $H || c$ at several temperatures. Small hysteresis is observed in $M$ and $\rho_{yx}$ at $2$ K, as seen as the difference between closed and open circles.}
\label{fig2}
\end{center}
\end{figure}

In Fig. 2(a), we show the magnetoresistance below $90$ K. The magnetoresistance is as small as $2$\% even at $2$ K and $14$ T. Figure 2(b) shows the $H$($||c$) dependence of $M$ at several temperatures. The slope of $M$-$H$ curve increases with decreasing $T$, but $M$ does not saturate up to $14$ T even at $2$ K; the $M$ at $14$ T is about $0.35 {\rm \mu}_{B}$/f.u., which is far smaller than the full moment value ($\sim 0.9 {\rm \mu}_{B}$/f.u.) of Fe determined by neutron diffraction measurement \cite{yashiro}. Since the magnetization of Fe(l) ($M^{l}$) would not change largely below $T_{N}$, the increase of $M$ below $T_{N}$ should be attributed to the increase in the magnetization of Fe(i) ($M^{i}$). For the latter use in the analysis of the Hall resistivity, we estimated the $M^{l}$ value (assumed to be $T$-independent \cite{note-on-M^l}) by $M^{l} = \chi^{l}H$, where the $T$-independent susceptibility for Fe(l) spins ($\chi^{l}$) is obtained by the subtraction of the Fe(i) component $\propto 1/(T+T_{0})$ from $M$ below $T_{N}$ in Fig. 1(b). (Here, the fitted Curie-Weiss temperature $T_{0}$ is close to $T_{g}\sim 10$ K.) Thus, the obtained $M^{l}$ value below $T_{N}$ based on this assumption is $H$-linear by definition, and reaches, for example, $0.17 {\rm \mu}_{B}$/f.u. at $14$ T. The observed non-linear $H$-dependence of $M$ in the low-$T$ region is ascribed to $M^{i}$ ($\equiv M-M^{l}$ at every temperature below $120$ K). At $2$ K, the hysteresis is observed in the $M$-$H$ curve up to around $5$ T, perhaps reflecting the freezing of Fe(i) spins or micto-glass state.
\par 
 
Figure 2(c) shows the Hall resistivity ($\rho_{yx}$) measured in $H||c$. $\rho_{yx}$ is negative and almost linear with $H$ above $60$ K. The roughly estimated carrier density ($\equiv -1/R_{0}e$, where $R_{0}$ is normal Hall coefficient) and mobility ($\equiv |R_{0}|/\rho_{||}$) above $T_{N}$ are $1.2 \times 10^{23}$/cm$^3$ and $0.16$ cm$^{2}$/Vs, respectively. Since the magnitude of $\rho_{yx}$ increases with decreasing $T$ in proportional to $M$, the contribution from the anomalous Hall effect (AHE) as induced by the spin-orbit interaction seems to be dominant below $T_{N}$. Below $40$ K, $\rho_{yx}$ shows an almost linear $H$-dependence in the low-$H$ region, but tends to curve positively in the high-$H$ region. Below $20$ K, $\rho_{yx}$ increases toward a positive direction in high-$H$ region, and at $2$ K shows even a sign-reversal around $13$ T, while the magnetoresistance remains as small as a few \%. Judging from the $T$ region in which the anomaly takes place, the order of the Fe(i) spins affects the AHE, whereas no distinct change in the $H$ dependence of $M$ nor $\rho_{||}$ is observed. 
\par

\begin{figure}[hbtp]
\begin{center}
\includegraphics[width=8cm]{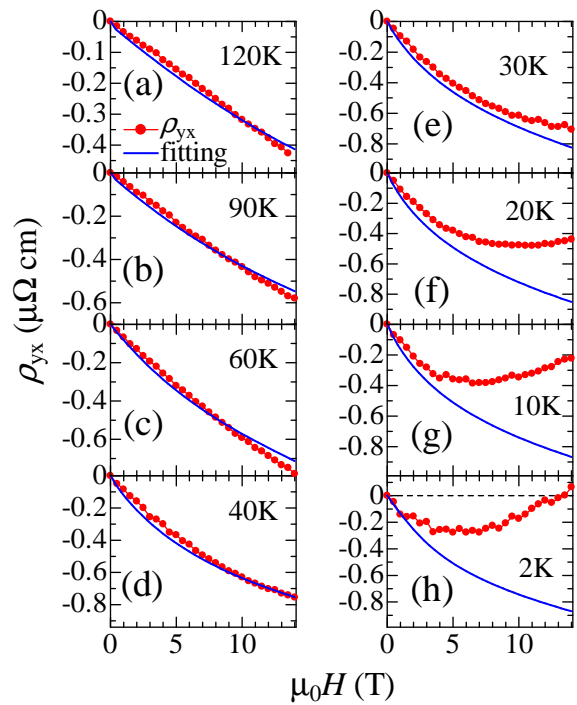}
\caption{(Color online.) (a)-(h) Fitting (solid lines) of the Hall resistivity $\rho_{yx}$ with the relation of $\rho_{yx} = R_{0}(\mu_{0}H) + R_{s}^{i}M^{i} +R_{s}^{l}M^{l}$ below $120$ K. Fitting parameters $R_{0}$, $R_{s}^{i}$, and $R_{s}^{l}$ are assumed to be $T$-independent (see text).}
\label{fig3}
\end{center}
\end{figure}

The Hall effect in a magnet is generally described by the empirical formula $\rho_{yx} = R_{0}(\mu_{0}H) + R_{s}M$, where $R_{0}$ and $R_{s}$ are normal and anomalous Hall coefficients, respectively. For the present material, the contributions from the moments of Fe(l) and Fe(i) should be distinguished, {\it i.e.}  $\rho_{yx} = R_{0}(\mu_{0}H) + R_{s}^{i}M^{i} +R_{s}^{l}M^{l}$ ($R_{s}^{i}$ and $R_{s}^{l}$ are $R_{s}$ for Fe(i) and Fe(l), respectively.). As shown in Figs. 3(a)-(d), we can fit nicely the experimental data above $40$ K using this relation with constant values of $R_{0}$, $R_{s}^{l}$, and $R_{s}^{i}$. The obtained values of $R_{0}$, $R_{s}^{l}$, and $R_{s}^{i}$ are $-4.9\times 10^{-9}$ ${\rm \Omega}$cm/T, $-1.07 \times 10^{-6}$ ${\rm \Omega}$cm/($\mu_{B}$/f.u.), and $-3.59 \times 10^{-6}$ ${\rm \Omega}$cm/($\mu_{B}$/f.u.), respectively. The good agreement between $\rho_{yx}$ and the fitting curve with use of these parameters above $40$ K (Figs. 3(a)-(d)) indicates that the Hall effect in the triangular lattice with an in-plane $120^\circ$-spin order can be understood within the framework of the conventional AHE. Below $30$ K, by contrast, $\rho_{yx}$ deviates from the fitting curve as shown in Figs. 3(e)-(h). (At $2$ K, we used the average values of $M^{i}$ and $\rho_{yx}$ in the hysteresis region for the fitting.) To explain such a large deviation from the conventional fit, one may consider the possibility that the $R_{s}^{i}$ value changes with the ordering of the Fe(i) spins around $10$ K. However, a dramatic change of $R_{s}^{i}$, including a sign-reversal in $H$ and steep $H$- and $T$-dependences, would be necessary for the explanation of the $H$ dependence of $\rho_{yx}$; this is highly unlikely. Moreover, $R_{0}$ and $R_{s}^{l}$ show negative signs above $40$ K and these values could be hardly changed by the ordering of the Fe(i) spins. It was confirmed that the different estimation of $M^l$ and $M^i$ least affects the emergent deviation below $30$ K, although does slightly the accuracy of the fitting curve above $40$ K. Thus, the difference between $\rho_{yx}$ and the fitting curve below $30$ K is not explained by the conventional formula and regarded as an unconventional term, which we assign here to the spin chirality mechanism. 

\par

We estimate the chirality-driven $\rho_{yx}$ ($\rho_{yx}^{\chi}$) as the deviation from the conventional normal and anomalous terms (see Fig. 3) and plot its $H$ dependence below $40$ K in Fig. 4(a). $\rho_{yx}^{\chi}$ at $14$ T is almost zero at $40$ K, but below $30$ K increases with decreasing $T$, as shown in the inset to Fig. 4(a). The onset $T$ for the $H$ ($14$ T)-induced spin chirality is tangibly higher than the transition $T$ of the Fe(i) spins ($\sim 10$ K), which indicates the spin glass state is not directly related with the spin chiraliy mechanism. At $2$ K, $\sigma_{xy}^{\chi} = \rho_{yx}^{\chi}/ \rho_{||}^{2}$ is $\sim 15$ ${\rm \Omega}^{-1}$cm$^{-1}$ at $14$ T. This value is as large as $\sigma_{xy}^{\chi}$ observed in Nd$_{2}$Mo$_{2}$O$_{7}$, in which the scalar spin chirality plays a dominant role due to the tilting ($\sim 4^\circ$) of the Mo spins from the ferromagnetic alignment \cite{Taguchi}. The $H$-dependence of $\rho_{yx}^{\chi}$ is approximately linear or slightly superlinear. 
\par

\begin{figure}[hbtp]
\begin{center}
\includegraphics[width=8cm]{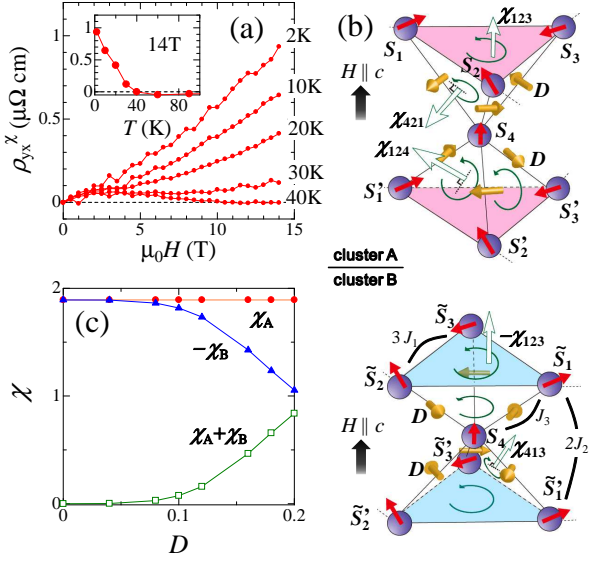}
\caption{(Color online.) (a) Magnetic-field ($H$) dependence of the chirality-driven Hall resistivity, $\rho_{yx}^{\chi}$, estimated from Fig. 3. Inset shows temperature ($T$) dependence of $\rho_{yx}^{\chi}$ at 14 T. (b) Schematic views of two types of the heptamer spin-clusters, A and B, consisting of one Fe(i) spin and two Fe(l)-spin triads above and below it under $H||c$ at low temperatures. Here the Fe(i) spins are assumed to be well $H$-polarized. Vectors normal to the faces as described by (green) open arrows denote directions of the scalar spin chirality from the respective triangle faces of the heptamer. The (yellow) arrows indicate directions of the DM vectors $\bm D_{ij}$ on the nearest-neighbor Fe(l)-Fe(i) bonds. (c)  Scalar spin chiralities $\chi_{\rm A}$ and $\chi_{\rm B}$ of the cluster A and B, and their sum $\chi_{\rm A}+\chi_{\rm B}$ as functions of $D$ (normalized by the in-plane interaction $J_{1}$) obtained by a numerical calculation (see text).}
\label{fig4}
\end{center}
\end{figure}

In the relatively-high-$T$ region below $T_N$, the Fe(i) spins are 
almost disordered due to thermal agitation, and the spin order 
occurs only as the in-plane 120$^{\circ}$-spin structure on the Fe(l) 
triangular lattices as shown in Fig. 1(d). In this situation, the net 
scalar spin chirality should cancel out in $\bm k$-space as suggested 
theoretically~\cite{Martin, Akagi}, and, thus, no additional Hall effect 
is expected, being consistent with our observation. 
On the other hand, in the low-$T$ region where the Fe(i) spins 
tend to be polarized by external $H$, there appear some heptamer 
spin-clusters in which one Fe(i) spin couples to two Fe(l)-triangles 
with 120$^\circ$-spin orders located above and below the Fe(i). 
Because of the alternate stacking of Fe(i) and Sb atoms, there are two 
kinds of the heptamer clusters with the equal population, A and B, as illustrated in Fig. 4(b). 
The total scalar spin chirality $\chi_{\rm A}$ ($\chi_{\rm B}$) 
for the heptamer cluster A (B) can be calculated as a sum of vectors 
normal to the respective triangular faces as exemplified by (green) 
open arrows, whose lengths are $\chi_{ijk}$. When only the symmetric 
exchange ($J_{ij} {\bm S}_{i} \cdot {\bm S}_{j}$) works among the Fe spins, the relation 
$\bm S_m$=$\bm S'_m$=$\tilde{\bm S}_m$=$\tilde{\bm S}'_m$ 
holds for $m=1$, $2$, and $3$ even under $H$$\parallel$$c$. 
Consequently $\chi_{\rm A}$ coincides with $-\chi_{\rm B}$, 
both of which are given by
$(2/3)(\chi_{124}+\chi_{234}+\chi_{314})+2\chi_{123}$.
This is because the cluster B is a mirror image of the cluster A, and
each face of the heptamer gives an opposite contribution between
A and B. For example, each of the basal and top Fe(l) triads gives 
$\chi_{123}$ for the cluster A, while $-\chi_{123}$ for the cluster B.
The perfect cancelation of $\chi_{\rm A}$ and $\chi_{\rm B}$ indicates 
that we need more ingredients to account for the observed topological 
Hall effect (THE).

Incorporation of the DM interaction, 
$\bm D_{ij} \cdot (\bm S_i \times \bm S_j)$, changes the situation 
dramatically. The DM vectors $\bm D_{ij}$ are locally defined on the bonds connecting two spins, and become finite when the local inversion symmetry is absent. Because of the absence of 
inversion symmetry on the Fe(i)-Fe(l) bonds, there are finite DM vectors 
$\bm D_{ij}$ whose directions are illustrated by (yellow) arrows in Fig. 4(b).
 The DM interaction causes canting of the Fe(l) spins and thus modifies 
the spin structure. Since the vectors $\bm D_{ij}$ on the upper Fe(i)-Fe(l) 
bonds are opposite to those on the lower Fe(i)-Fe(l) bonds in each heptamer, 
the spin canting becomes inequivalent between the basal and top triangle 
planes, i.e., $\bm S_m \neq \bm S'_m$ and 
$\tilde{\bm S}_m \neq \tilde{\bm S}'_m$. 
Moreover the Fe(l) spins cant in a different manner between the 
clusters A and B. For the Fe(l) triads of the cluster A, directions of 
the three Fe(l) spins are modified while keeping the original chirality for
the case without DM interaction, whereas for those of the cluster B the 
spin canting alters the chirality value. Owing to the inequivalent 
modifications of the spin structure, the two contributions $\chi_{\rm A}$ and 
$\chi_{\rm B}$ no longer cancel out, and the net scalar spin chirality 
proportional to $\chi_{\rm A}+\chi_{\rm B}$ becomes nonzero, which 
works as an origin of the observed THE. 

The above discussion is confirmed by numerical simulations of a 
classical Heisenberg model on the heptamer spin clusters. In this model, 
the spins are treated as classical vectors whose norms are set to be unity. 
The model contains the symmetric exchange interactions, DM interaction, 
magnetic anisotropy, and Zeeman coupling. For the symmetric exchange 
interactions, we consider antiferromagnetic $3J_1$ between the 
in-plane Fe(l)-Fe(l) spin pairs, ferromagnetic $2J_2$ between the 
out-of-plane Fe(l)-Fe(l) spin pairs, and weak antiferromagnetic $J_3
$ between the Fe(i)-Fe(l) spin pairs [see Fig. 4(b)]. 
Since the Fe(l) spins in the 120$^{\circ}$ order direct along the in-plane bonds at $H$=0, we 
includes the following anisotropy term, $-A \sum_{i}[(\bm S_i \cdot \bm 
l)^2+(\bm S_i \cdot \bm m)^2+(\bm S_i \cdot \bm n)^2]$ where $\bm l$ and 
$\bm m$ are the Bravis vectors of the triangular lattice and 
$\bm n$=$\bm l$+$\bm m$. 
The Zeeman couping is given by $-H \sum_{i} S_{zi}$. For the parameter
values to simulate the low-$T$ case, we take $J_1$=1, $J_2$=$-$0.5, $J_3$=0.2, $A$=0.2, and $H$=0.1.
 We numerically search the lowest-energy spin cofigurations, and calculate 
the scalar spin chiralities $\chi_{\rm A}$ and $\chi_{\rm B}$ as functions 
of strength of the DM vector $|\bm D_{ij}|$=$D$ as shown in Fig.4(c). In the 
absence of the DM interaction ($D$=0), $\chi_{\rm A}$ and $-\chi_{\rm B}$ are equivalent. As $D$ increases, $-\chi_{\rm B}$ 
decreases whereas $\chi_{\rm A}$ does not change, which results in 
finite $\chi_{\rm A}+\chi_{\rm B}$. In reality, for $D$ to be effective to generate the spin chirality (empirically, $D/J=0.1 \sim 0.2$), the averaged moment value of Fe(i) spin and the canting angle of the in-plane Fe(l) spin moments by DM interaction are essential; the former is mainly determined by magnetic field, while the latter by $D/J$. This may explain the low temperature ($30$ K) and the nearly $H$-linear evolutions of THE as observed. 
\par

In conclusion, we have investigated the Hall effect in the triangular-lattice Fe$_{1.3}$Sb under $H$ applied perpendicular to the basal plane. In the relatively-high-$T$ region of the $120^\circ$ spin ordered structure on the triangular lattice, the Hall resistivity is negative in sign and well described as composed of the conventional normal and anomalous components. In the low-$T$ region below $30$ K, however, the significant positive contribution is observed to be added to the Hall resistivity. This novel term is explained in terms of the scalar spin chirality which originates from the spin modulation by DM interaction in the spin heptamer clusters formed around the respective Fe(i) sites.

\par

We thank S. Miyahara, N. Furukawa, N. Nagaosa, G. Tatara, and Y. Onose for fruitful discussions. This work was in part supported by the Grant-in-Aid for Scientific Research (Grant No. 20340086.) and JSPJ Fellows, and by the Funding Program for World-Leading Innovate R\&D on Science and Technology (FIRST program) on ``Quantum Science on Strong Correlation".

\newpage


\begin{thebibliography}{99}
\bibitem{chirality-note}Even in such a case, the HE induced by the scalar spin chirality may emerge in materials with the crystal structure including inequivalent multiple loops in a unit cell \cite{skyrmion, Ohgushi}. The spin-chirality induced HE in pyrochlore magnets correspond to this category.
\bibitem{note-on-M^l}This assumption is based on the generic feature of the $T$-independent {\it transverse} susceptibility for the antiferromagnetic state with the strong anisotropy.
\end{thebibliography}
\end{document}